\documentclass[twocolumn]{aastex631}

\usepackage{amsmath}
\usepackage{enumitem}

\begin{document}

\title{Lopsided and Bulging Distribution of Satellites around Paired Halos. II. 3D Analysis and Dependence on Projection and Selection Effects.}

\correspondingauthor{Qinglin Ma}
\email{maql21@mails.tsinghua.edu.cn}
\correspondingauthor{Cheng Li}
\email{cli2015@tsinghua.edu.cn}

\author[0009-0000-5148-9457]{Qinglin Ma}
\affiliation{Department of Astronomy, Tsinghua University, Beijing 100084, China}

\author[0000-0002-8711-8970]{Cheng Li}
\affiliation{Department of Astronomy, Tsinghua University, Beijing 100084, China}

\author{Yanhan Guo}
\affiliation{Department of Astronomy, Tsinghua University, Beijing 100084, China}



\begin{abstract}
We use the Illustris-TNG simulation to investigate the anisotropic distribution of subhalos in/around dark matter halo pairs. We measure the position angle ($\theta$) of each subhalo by the angle between the line connecting it to the nearest host halo and the line connecting the paired halos, and examine $P(\cos\theta)$ (the distribution of $\cos\theta$ of all subhalos) for halo pairs with various separations ($d_{\text{sep}}$), primary halo masses ($M_p$) and secondary-to-primary halo mass ratios ($M_s/M_p$). We find that $P(\cos\theta)$ generally exhibits a combined result of two distinct features: the “bulging” distribution characterized by an overabundance along the pairwise direction, and the “lopsided” distribution showing an overabundance in the region between the paired halos. The bulging signal is stronger for halo pairs with larger $d_{\text{sep}}$ and smaller $M_p$, while the lopsidedness strengthens as $M_p$ increases. Both signals depend weakly on $M_s/M_p$, and are primarily contributed by subhalos that are relatively distant from host halos. Remarkably, these measurements can be broadly reproduced by the overlap effect, provided the spatial alignment of halos is properly taken into account. Our findings suggest distinct origins: lopsidedness arises from simple halo overlap, while bulging reflects alignment with large-scale filaments. We examine the impact of projection and selection effects by conducting the same analysis in two dimensions and in a mock catalog that replicates the selection effects of the SDSS galaxy sample. We find that the 3D-to-2D projection significantly suppresses the bulging distribution, with particularly strong effects at large $d_{\text{sep}}$, small $M_p$, and large $M_s/M_p$. 
\end{abstract}



\section{Introduction} \label{sec:intro}

Matter distribution in the universe is not uniform on scales smaller than a few hundred Mpc. Instead, it presents a complex pattern known as the cosmic web \citep{1996Bond}, comprising interconnected {\em filaments} of galaxies and gas, dense galaxy clusters and massive dark matter halos ({\em nodes}), vast empty regions ({\em voids}), and flattened {\em walls}. The cosmic web  serves as a manifestation of the intricate large-scale structure (LSS) of the universe, shaped by gravitational collapse due to dark matter and negative pressure due to dark energy, in addition to cosmic expansion. Therefore, understanding the cosmic web is crucial for shedding light on galaxy formation, dark matter properties, and cosmological models. 

In particular, the cosmic web is believed to play vital roles in regulating the orientations of both dark matter halos and galaxies, as well as the anisotropic distribution of their surrounding subhalos and satellite galaxies. 
For instance, the tidal field generated by the cosmic web is capable of distorting the collapsing protogalactic halos (\citealt{1969Peebles}; \citealt{1970Doroshkevich}; \citealt{1984White}), contributing to the anisotropic accretion of satellites (\citealt{2004Knebe}; \citealt{2004Aubert}; \citealt{2005Libeskind}; \citealt{2005Zentner}; \citealt{2014Wang}; \citealt{2018Shao}), and subsequently shaping the alignment between halo and galaxy orientations and the large-scale filamentary structure (\citealt{2009Faltenbacher}; \citealt{2009Zhang}; \citealt{2011Paz}; \citealt{2012Libeskind}; \citealt{2012Schneider}; \citealt{2013Li}; \citealt{2014Forero}; \citealt{2015Libeskind}; \citealt{2015Kang}; \citealt{2018Ganeshaiah}).

The anisotropic distribution of satellites has been observed around both isolated galaxies in our universe and dark matter halos in numerical simulations. This anisotropy manifests as two distinct features: a lopsided distribution with an overabundance of satellites located on one side of their host halos or central galaxies (\citealt{2020Brainerd}; \citealt{2021Wang}; \citealt{2023Samuels}; \citealt{2024Heesters}), and a bulging distribution, characterized by the preferential location of satellites along the major axis of the host halos or central galaxies (\citealt{2008Faltenbacher}; \citealt{2014Libeskind}; \citealt{2015TempelB}; \citealt{2022Gu}). It is well established that both the bulging and lopsided satellite distributions around isolated halos or galaxies originate from the large-scale tidal field and the anisotropic accretion of galaxies and subhalos (e.g., \citealt{2004Knebe}; \citealt{2004Aubert}; \citealt{2005Libeskind}; \citealt{2005Zentner}; \citealt{2015Kang}; \citealt{2024Liu}).

The anisotropic distribution of satellites has also been observed around pairs of galaxies or dark matter halos. For instance, a lopsided distribution has been reported for the satellite galaxies in the Local Group, which are preferentially located in the region between the Milky Way (MW) and M31 (\citealt{2006McConnachie}; \citealt{2013Conn}; \citealt{2013Ibata}). Furthermore, \citet{2016Libeskind} examined the satellites around a sample of galaxy pairs similar to the MW-M31 system, statistically detecting a similarly anisotropic distribution. This anisotropy is characterized not only by a lopsided feature with elongated filamentary structures, but also by a bulging feature, where more satellites are observed along the line connecting the galaxy pairs than those seen perpendicular to it. In line with this, \citet{2017Pawlowski} conducted a search for analogous anisotropy in the Local Group analogs within cosmological simulations.

In a parallel paper (\citealt{2025Guo}; hereafter Paper I), we have extended the study of lopsided and bulging distributions of satellites around halo pairs, by selecting a sample of paired central galaxies from the Sloan Digitak Sky Survey \citep[SDSS;][]{2000York} and examining the lopsidedness and bulging signals as function of host halo mass, halo mass ratio and pair separation. In addition, we have constructed a mock catalog that has the same selection effects as the observational sample, based on the Illustris-TNG300 simulation. By incorporating the orientation of individual galaxies and the alignment of galaxy/hao orientation with halo pair orientation, our mock catalog successfully reproduces the lopsided and bulging distributions of satellites across halo pairs of various properties. Our results demonstrate that the angular distribution of satellites is a natural phenomenon of the $\Lambda$CDM model, eliminating the need to introduce other mechanisms such as the local gravitational effects proposed by \citet{2016Libeskind}. 

In this work, we perform a three-dimensional (3D) analysis of the lopsided and bulging distribution of satellites around pairs of dark halos, examining the potential biases in the observational measurements caused by projection and selection effects in redshift surveys. For this purpose, we measure the angular distribution of subhalos in and around halo pairs with varying separations, halo masses, and halo mass ratios, similar to our approach in Paper I. In particular, we obtain measurements in three-dimensional space using the full simulation box, in addition to those in a projected two-dimensional plane and in redshift space using a mock catalog that mimics the selection effects of SDSS-like surveys. By comparing the results from these three different conditions, we aim to discern the influence of projection and selection effects on the angular distribution of satellites, as observed in previous studies. As we will demonstrate, the projection effect significantly weakens the bulging distribution of satellites in halo pairs but has a negligible effect on their lopsided distribution. The three-dimensional analysis confirms that the scenario revealed by Paper I and previous studies based on observational samples remains broadly valid.

In what follows, we first present 3D analyses of the satellite distribution around halo pairs and overlapping halos in \autoref{sec:3danalysis}. We then investigate the influence of projection and selection effects in \autoref{sec:projection}. We discuss in \autoref{sec:discussion} and conclude in \autoref{sec:conclusions}.

\section{3D Analysis}
\label{sec:3danalysis}

\subsection{Satellite distribution around halo pairs}
\label{sec:3_2}

Througout this work, we make use of the \textit{Next Generation Illustris} dark matter only simulation (IllustrisTNG-ODM; \citealt{2018Marinacci}; \citealt{2018Naiman}; \citealt{2018Nelson}; \citealt{2018Springel}; \citealt{2018Pillepich}), which contains $2500^{3}$ dark matter particles, each with a mass of $7 \times 10^{7}h^{-1}\mathrm{M}_{\odot}$, in a peroid box of size $L = 205 h^{-1} \mathrm{Mpc}$ on a side. The adopted cosmology parameters are: $\Omega_m = 0.3089$, $\Omega_b = 0.0486$, $\Omega_{\Lambda} = 0.6911$, $h = 0.6774$, $\sigma_8 = 0.8159$, and $n_s = 0.9667$. Dark matter halos and subhalos are identified using the friends-of-friends \citep[FoF;][]{Davis1985} and the {\tt subfind} \citep{2001Springel} algorithms, respectively. Throughout this work, we focus our analysis on the local universe, thus using only the simulation snapshot at $z=0$.

\begin{figure}[htbp]
\centering
	\includegraphics[width=0.95\columnwidth]{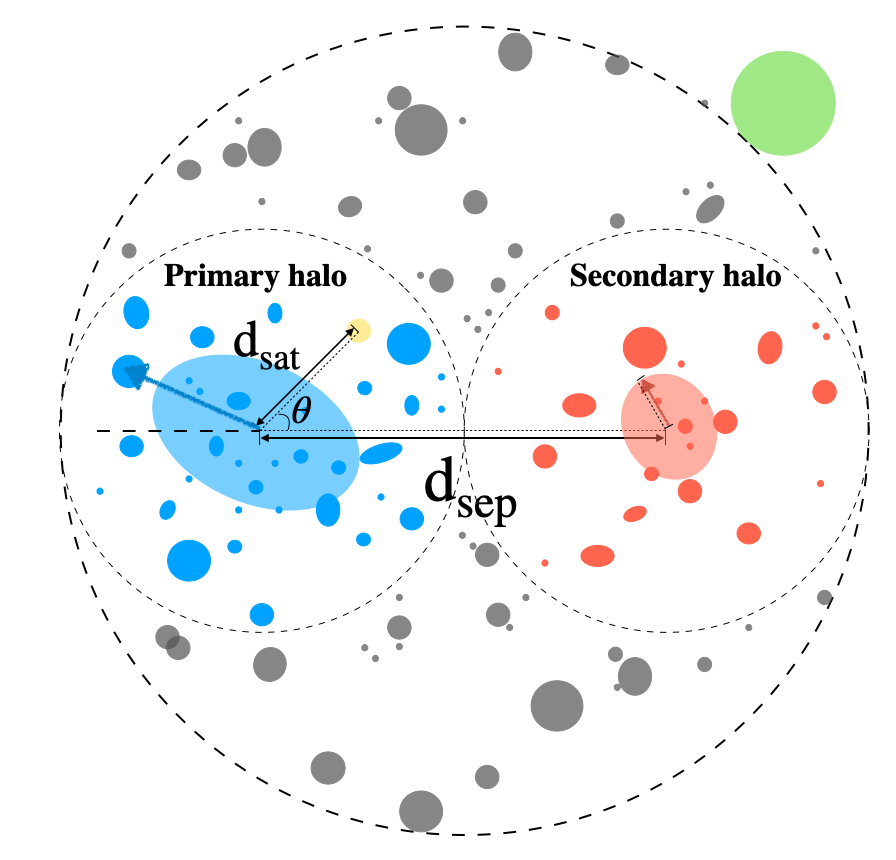}
    \caption{The schematic of a typical halo pair is illustrated, where the more massive halo (primary, depicted in blue) and the less massive halo (secondary, depicted in red) are separated by a distance of $d_{\mathrm{sep}}$. Subhalos located within a radius of $0.5 d_{\mathrm{sep}}$ from the two halos are considered ``satellites'', represented by red and blue ellipticals enclosed by thin dashed circles. For a given satellite at a distance of $d_{\text{sat}}$ from its host halo, the position angle $\theta$ is defined as the angle between the line connecting the satellite to its halo and the line connecting the paired halos. The large dashed circle, which is centered at the midpoint of the pair connection line and has a radius of $d_{\mathrm{sep}}$, represents the exclusion boundary, within which there must be no third halo that is more massive than half of the primary halo mass. See the text for further details.}
    \label{fig:s_diagram}
\end{figure}

From the simulation, we select halo pairs in three dimensions, requiring that the involved halos in each pair are separated by a distance in the range of $0 < d_{\mathrm{sep}}< 2 h^{-1}\mathrm{Mpc}$ and have masses in the range of $10^{11.6} h^{-1}\mathrm{M}_{\odot} < \mathrm{M} < 10^{15.6} h^{-1}\mathrm{M}_{\odot}$. The halo mass is defined as the total dark matter mass within the virial radius $R_{\text{vir}}$, within which the average matter density is 102 times the critical density of the universe. Additionally, there must be no third halo within a radius equal to the pair separation $d_{\mathrm{sep}}$ from the midpoint of the line connecting the two halos. For a given pair, the more massive halo is referred to as the {\em primary halo}, while the less massive one is referred to as the {\em secondary halo}. These restrictions give rise to a sample of 25,139 halo pairs at $z=0$. We note that our selection criteria are similar to those adopted in \citet{2019Gong}, but we have adopted wider ranges for pair separation and halo mass. This allows us to extend the analysis to more general cases beyond those resembling the Local Group. 

\autoref{fig:s_diagram} presents a schematic diagram of the halo pairs. Following \citet{2019Gong}, we consider all the subhalos located within $d_{\mathrm{sep}}/2$ from the nearer halo when measuring the satellite distribution for a halo pair. For the range of $d_{\mathrm{sep}}$ considered, we establish that $d_{\mathrm{sep}}/2 > R_{p,\mathrm{vir}}$ (the virial radius of the primary halo) holds for nearly all halo pairs. This criterion ensures the representative inclusion of subhalos residing within the virial radii of both halos. Furthermore, our satellite definition encompasses subhalos located beyond these virial radii, as visually indicated by the blue and red ellipses in the figure. These external subhalos are significant as they probe the connection between the halo pairs and the surrounding large-scale structure. For each subhalo in consideration, we measure its positional angle $\theta$, defined as the angle between its position vector with respect to the nearer halo and the line connecting the two halos. Defined this way, angles in the range of $0 \leq \theta \leq 90^{\circ}$ (or equivalently $0 \leq \cos\theta \leq 1$) correspond to subhalos located in the internal region between the two halos, while angles in the range of $90^{\circ} < \theta \leq 180^{\circ}$ (or equivalently $-1 \leq \cos\theta < 0$) correspond to those located on the outer sides.

\begin{figure*}
   \centering
   \includegraphics[width=0.8\textwidth]{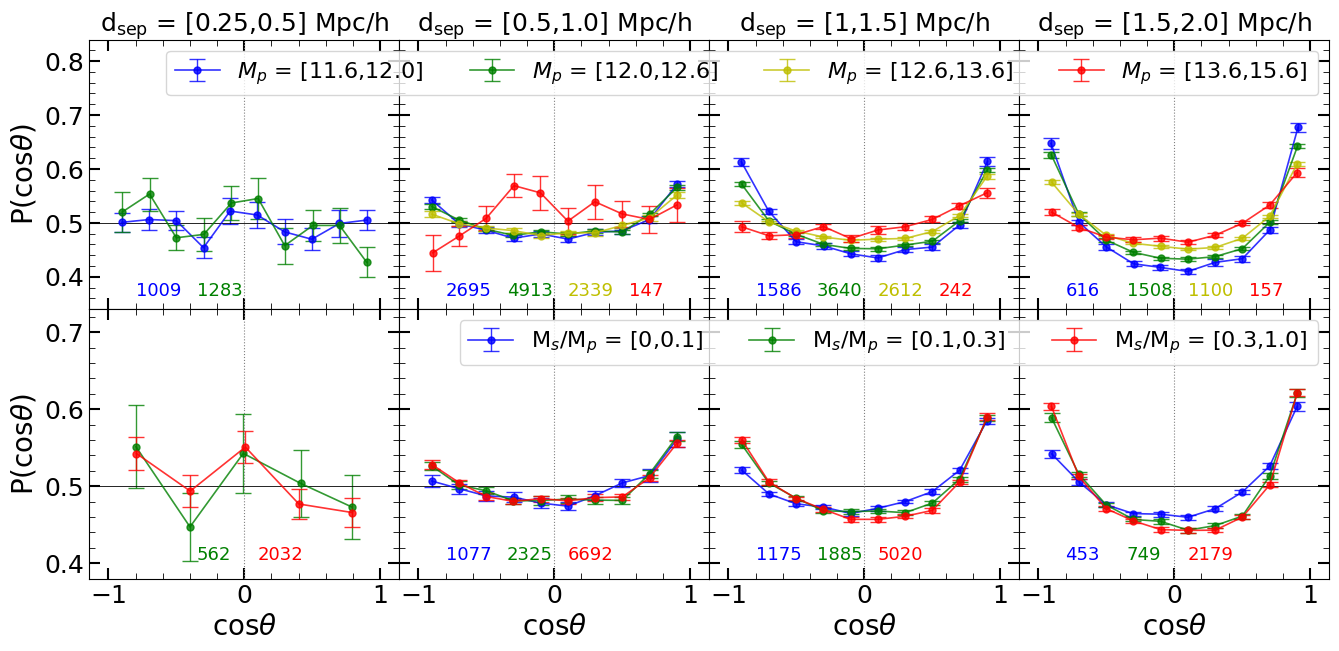}
   \caption{Satellite distribution in 3D as measured by $P(\cos \theta)$ is shown for halo pairs of varying separation d$_{\mathrm{sep}}$, primary halo mass M$_{\mathrm{p}}$, and secondary-to-primary halo mass ratio M$_{\mathrm{s}}$/M$_{\mathrm{p}}$, as indicated. The horizontal line indicates the uniform distribution. The numbers listed at the bottom are the sample size of the corresponding subsamples.}
   \label{fig:3d_dsep_only_pair}
\end{figure*}

\begin{figure*}
   \centering
   \includegraphics[width=0.8\textwidth]{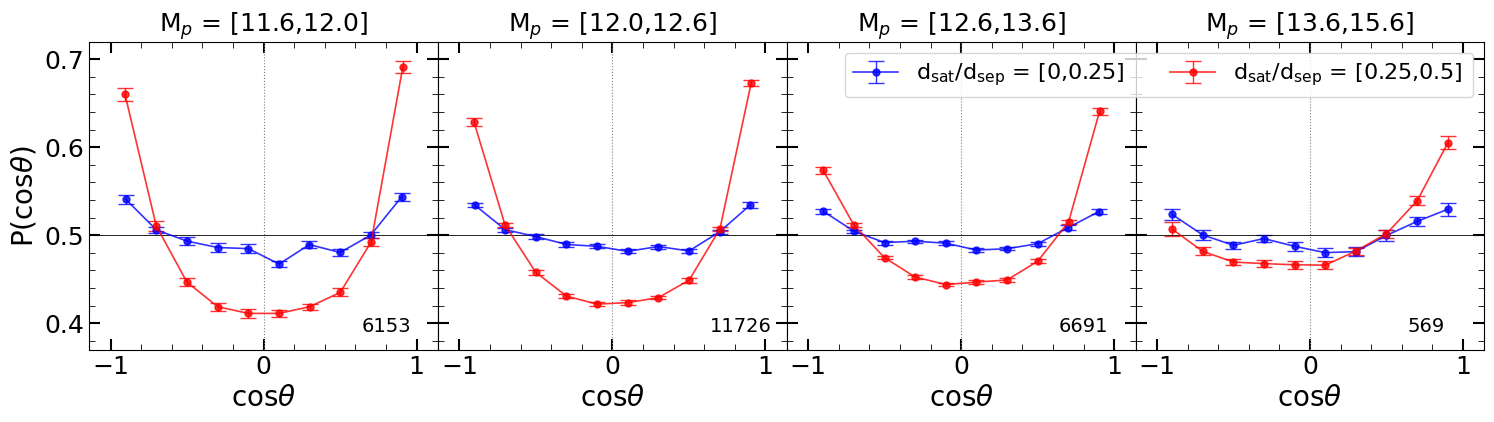}
   \caption{Satellite distribution $P(\cos \theta)$ is shown for halo pairs of different primary halo masses, as well as different satellite-to-host distance for given halo mass, as indicated. Those two subsamples have the same number of pairs, as listed at the bottom. }
   \label{fig:3d_dsat_only_pair}
\end{figure*}

\autoref{fig:3d_dsep_only_pair} displays the distribution of $\cos\theta$ for halo pairs with varying primary halo masses ($M_p$), pair separations ($d_{\text{sep}}$), and secondary-to-primary halo mass ratios ($M_s/M_p$). Additionally, \autoref{fig:3d_dsat_only_pair} compares the $\cos\theta$ distribution for subhalos at different distances from their host halo, as measured for halo pairs in different primary halo mass bins. In both figures and throughout the paper, errors of $P(\cos\theta)$ are estimated with the technique of bootstrap resampling, by randomly resampling 50 times the actual halo pairs or the overlapping halos in a given sample. Overall, $\cos\theta$ presents a non-uniform distribution for halo pairs with separations $d_{\text{sep}} > 0.5~h^{-1} \text{Mpc}$, showing a minimum of $P(\cos\theta)$ around $\cos\theta = 0$ and increased values towards both $\cos\theta = 1$ and $\cos\theta = -1$. This indicates a ``bulging'' distribution of satellites around the halo pairs, characterized by an overabundance of satellites along the connection line of the halo pairs, both inside and outside the pairs. The bulging signal is stronger for  halo pairs that are more widely separated and involve less massive halos, but it exhibits a rather weak dependence on the secondary-to-primary halo mass ratio. As shown by \autoref{fig:3d_dsat_only_pair}, the bulging signal at a given halo mass is primarily contributed by subhalos that are relatively distant from their host halos. 

In addition, lopsidedness is also observed for halo pairs with $d_{\text{sep}} > 0.5~h^{-1} \text{Mpc}$, indicating that satellites tend to be preferentially located inside the halo pairs ($\cos\theta > 0$) rather than outside ($\cos\theta < 0$). This signal strengthens as the halo mass increases, but it shows weak dependence on the pair separation. Similar to the bulging distribution, the lopsided distribution also exhibits a weak dependence on the halo mass ratio and is primarily attributed to more distant subhalos.

In contrast, the subhalos around close halo pairs with $d_{\text{sep}} < 0.5~h^{-1} \text{Mpc}$ appear to exhibit relatively uniform distributions, or even slightly higher abundances on the outer sides, though this is associated with large uncertainties. It is important to note that these close pairs are dominated by relatively low-mass halos and consist of halos with comparable masses. As a result, they are likely merging systems and dynamically unstable, leading to different processes regulating the satellite distribution around these pairs compared to the other halo pairs in our sample.

\begin{figure*}
  \centering
  \includegraphics[width=0.8\textwidth]{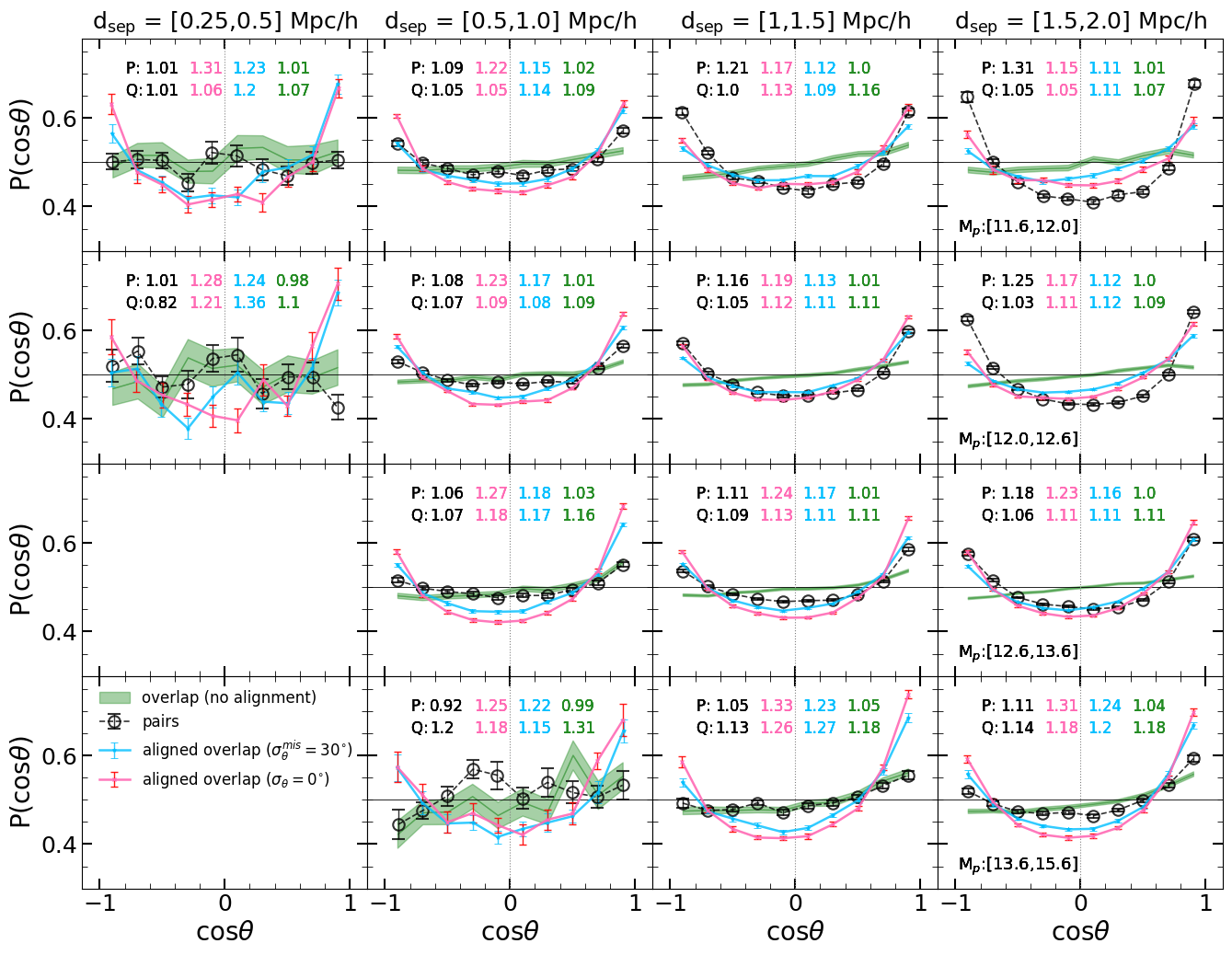}
  \caption{Satellite distribution $P(\cos\theta)$ for halo pairs of different pair separations (panels from left to right) and primary halo masses (panels from top to bottom), as indicated. In each panel, different symbols/lines/colors represent the results of different samples, including the actual halo pairs (black circles), the overlap sample without aligment (green), and the aligned overlap samples with a misalignment angle of zero (magenta) or 30$^\circ$ (cyan). The $P$ and $Q$ parameters quantifying the bulging and the lopsidedness in the satellite distribution are indicated in each panel, with the colors corresponding to the different samples.}
  \label{fig:3d_mass_dsep}
\end{figure*}

It might be questioned whether the relatively wide ranges of $d_{\text{sep}}$, $M_p$, and $M_s/M_p$ have introduced biases into our results, given the potential dependence of $d_{\text{sep}}$ on $M_p$ and $M_s/M_p$. To address this, we have examined the distribution of halo pairs in our sample on the $d_{\text{sep}}$ versus $M_p$ and $d_{\text{sep}}$ versus $M_s/M_p$ planes. Overall, $d_{\text{sep}}$ exhibits weak positive correlations with $M_p$ and weak negative correlations with $M_s/M_p$. However, when restricted to individual subsamples, these correlations become considerably weaker, indicating a negligible influence on the results presented in this paper.

\subsection{Satellite distribution around overlapping halos}

One possible explanation for the lopsided and bulging distribution of satellites around paired halos is the so-called ``overlap effect'': an overabundance of satellites is naturally found between two halos when they are brought closer together, along with their surrounding satellite distribution. In previous studies, this hypothesis has been tested by comparing the satellite distribution around actual halo pairs with that around overlapping halos (\citealt{2016Libeskind}; \citealt{2017Pawlowski}; \citealt{2019Gong}). For each actual halo pair, two halos with masses similar to those of the halos in the pair are selected from the simulation, and along with their surroundings, they are placed at the same separation as the actual halo to form the corresponding overlapping halos. In practice, we require the overlapping halos to be closely matched in mass with the corresponding halos in the actual pairs, with a tolerance of $\Delta \log_{10} M_h \leq 0.1$, and to be separated at the same distance as the actual halo pair, with a tolerance of $\Delta d_{\text{sep}} \leq 0.5 ~h^{-1} \text{Mpc}$. For each of the actual pairs, this process is repeated 200 times, resulting in a sample of overlapping halos that is 200 times the size of the actual pair sample. 

\begin{figure*}[ht!]
   \centering
   \includegraphics[width=0.8\textwidth]{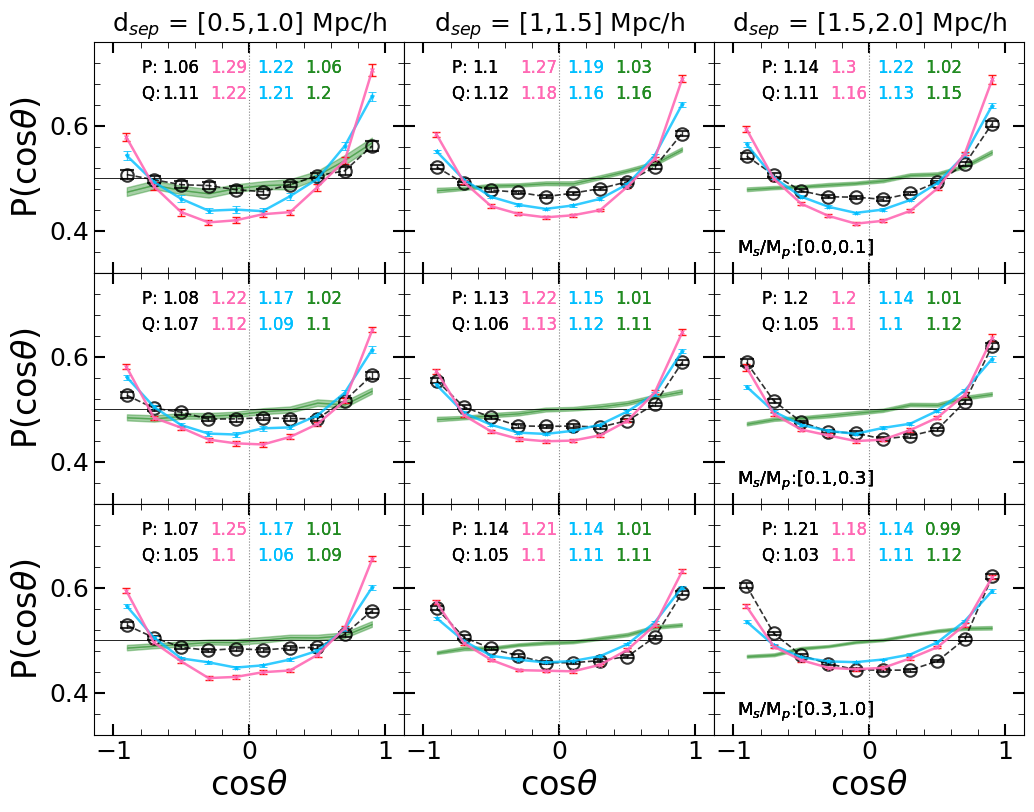}
   \caption{Satellite distribution $P(\cos\theta)$ for halo pairs of different pair separations and secondary-to-primary halo mass ratios, as indicated. The symbols and colors are the same as in~\autoref{fig:3d_mass_dsep}.}
   \label{fig:3d_dsep_mrate}
\end{figure*}

\begin{figure*}[ht!]
   \centering
   \includegraphics[width=0.8\textwidth]{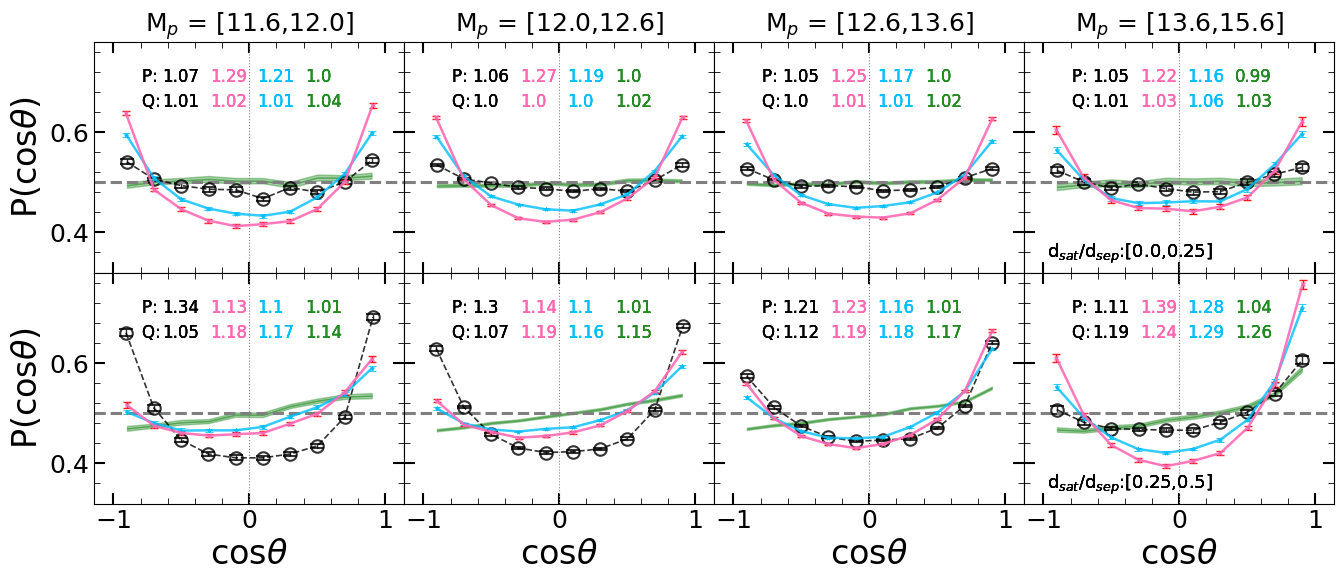}
   \caption{Satellite distribution $P(\cos\theta)$ for halo pairs of different primary halo masses and satellite-to-host distances, as indicated. The symbols and colors are the same as in~\autoref{fig:3d_mass_dsep}.}
   \label{fig:3d_mass_dsat_dsep}
\end{figure*}

Measurements of the satellite distribution around the overlapping halos at different separations are plotted as green shaded regions, categorized by different primary halo masses in \autoref{fig:3d_mass_dsep}, by various halo mass ratios in \autoref{fig:3d_dsep_mrate}, and by different satellite-to-host distances in \autoref{fig:3d_mass_dsat_dsep}. The corresponding measurements for the actual halos obtained earlier are plotted as black circles with error bars for comparison. Generally, the overlapping halos present no bulging signals but show only a lopsided distribution, with satellites being preferentially found between the two halos. Overall, the lopsidedness is relatively weak and matches that of the actual halo pairs only for close pairs with $d_{\text{sep}} < 0.5~h^{-1} \text{Mpc}$ or for those involving the most massive halos ($M_p > 10^{13.6}~h^{-1} \text{M}_\odot$), where the signals in actual halo pairs are also weak. This result indicates that the overlap effect alone is insufficient, and additional mechanisms are required to fully explain the satellite distribution around halo pairs.

When constructing the overlap sample, the two overlapping halos are randomly oriented, as done in previous studies. However, as pointed out in Paper I, dark matter halos are not oriented at random; rather, they are preferentially aligned with large-scale filamentary structures, a fact that has been well established both for dark halos in simulations and for galaxies and groups of galaxies in redshift surveys. Following Paper I, we incorporate the alignment of halos in the overlap sample, by aligning the orientation of the two overlapping halos with their connection line, which is known to be somewhat aligned with large-scale filaments (e.g. \citealt{2015Tempel}; \citealt{2018Mesa}; \citealt{2024Lamman}; \citealt{2024Rong}; \citealt{2024Sarkar}). To determine the orientation of halos, we use the dark matter particles within $R_{\text{vir}}$ to calculate the weighted inertia tensor for each halo~\citep{2006Allgood}. Assuming the halo can be modeled by a triaxial ellipsoid, we then determine the three axes of the halo by the eigenvectors of the inertia tensor, with the major axis given by the eigenvector corresponding to the largest eigenvalue. For each pair of overlapping halos in the overlap sample, we rotate the entire configuration of subhalo distribution around each of the two halos, ensuring that the major axes of the two halos are aligned with each other. 

The measurements of $P(\cos\theta)$ for the ``aligned overlap sample'' constructed in this manner are additionally plotted as magenta lines in \autoref{fig:3d_mass_dsep}, \autoref{fig:3d_dsep_mrate}, and \autoref{fig:3d_mass_dsat_dsep}. Furthermore, we randomly rotate the major axis of the overlapping halos to mimic the misalignment of halo orientations found in previous studies (e.g. \citealt{2009Faltenbacher}; \citealt{2009Okumura}). We adopt a Gaussian distribution for the misalignment angle, with a mean of zero and a width of $\sigma_{\text{mis}} = 30^{\circ}$. Here, $\sigma_{\text{mis}}$ is chosen to represent the average misalignment of different types of galaxies, which have been found to be $25^{\circ} - 35^{\circ}$ for red galaxies and $25^{\circ} - 65^{\circ}$ for blue galaxies (\citealt{2008Wang}; \citealt{2009Okumura}). The measurements of $P(\cos\theta)$ for this ``misaligned overlap sample'' are plotted as cyan lines in the same figures.

To enable quantitative comparisons, we define two parameters, $P$ and $Q$, to respectively quantify the strength of the bulging and lopsidedness signals. Specifically, $P$ is defined as the integral of $P(\cos\theta)$ over the ranges $0.5 < \cos \theta < 1.0$ and $-1.0 < \cos \theta < -0.5$, divided by the integral of $P(\cos\theta)$ over the range $-0.5 < \cos \theta < 0.5$. This parameter measures the relative abundance of satellites along the direction of the pair orientation compared to that perpendicular to it. Following \cite{2019Gong}, the $Q$ parameter is defined as the ratio of $P(\cos\theta)$ at $0.8 < \cos \theta < 1.0$ to that at $-1.0 < \cos \theta < -0.8$, thereby measuring the relative satellite abundance between the inner and outer sides of the halo pair. The $P$ and $Q$ parameters are calculated for all the $P(\cos\theta)$ measurements, as indicated in each panel of \autoref{fig:3d_mass_dsep}, \autoref{fig:3d_dsep_mrate}, and \autoref{fig:3d_mass_dsat_dsep}. 

As can be seen from the figures and according to the $P$ and $Q$ parameters, after the alignment of halos is included, the satellite distribution in all cases becomes strongly bulging, while the lopsidedness remains similar. The inclusion of misalignment only slightly weakens the bulging signal. Consequently, both the ``aligned'' and ``misaligned'' overlap samples can roughly replicate the behaviors of actual halo pairs in most cases, but they overpredict the bulging signal for close pairs or pairs with massive halos, where the original, randomly placed overlapping halos provide better agreement. This result implies that the lopsidedness signal and the bulging signal may arise from two distinct origins: the bulging relates to the alignment of halo orientation and subhalo distribution with large-scale filamentary structures, which is significant for relatively widely separated pairs of less massive halos, and the lopsidedness is simply the overlap effect, which applies to pairs of all cases but plays a dominant role only for pairs of massive halos or substantially close pairs of low-mass halos. We will revisit this topic and discuss it further in \autoref{sec:discussion_result}.

\section{Projection and selection effects}
\label{sec:projection}

\begin{figure*}[ht!]
    \centering
	\includegraphics[width=0.8\textwidth]{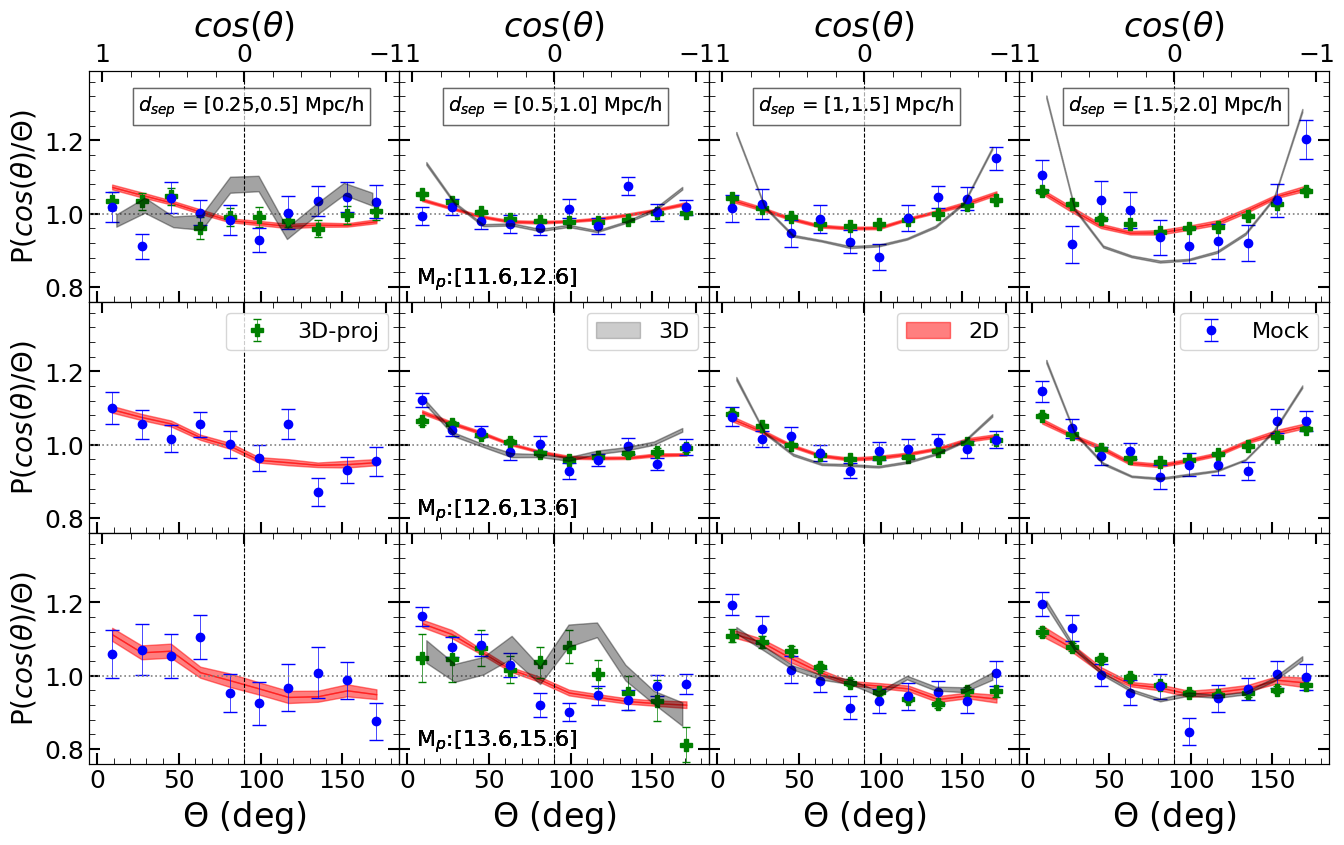}
    \caption{Satellite distribution $P(\Theta)$ obtained in 2D and from the mock catalog are shown in red shaded regions and blue dots, for halo pairs of different separations and primary halo masses, as indicated. For comparison, the 3D measurement $P(\cos\theta)$ and the 2D measurement of the halo pairs selected in 3D are plotted as grey shaded region and green dots, respectively.}
    \label{fig:3d_2d_mock_dms_pair}
\end{figure*}

\begin{figure*}[ht!]
    \centering
	\includegraphics[width=0.8\textwidth]{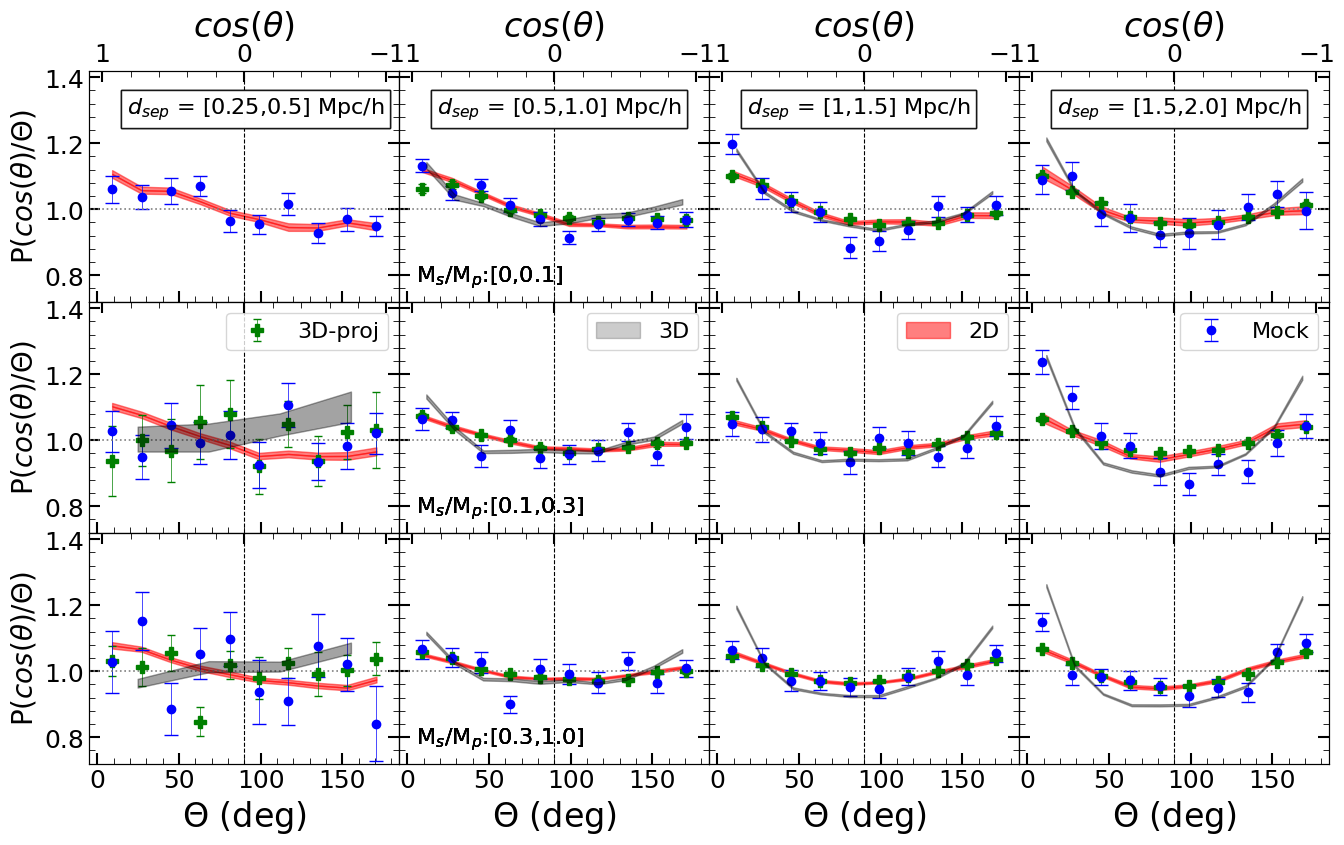}
    \caption{Same as the previous figure, but panels from top to bottom are for halo pairs of different secondary-to-primary halo mass ratios, as indicated.}
    \label{fig:3d_2d_mock_dmr_pair}
\end{figure*}

Observational studies of galaxy distribution rely on imaging and flux-limited samples, which suffer from projection and complex selection effects. It is crucial to consider these effects before properly interpreting the observational results. In this section, we examine the satellite distribution around halo pairs and overlapping halos, both in two dimensions by projecting the simulation box onto a 2D plane, and in redshift space by constructing mock catalogs that replicate the selection effects of real galaxy surveys.

\subsection{Satellite distribution around halo pairs}

For the 2D analysis, we take the $x-y$ plane of the IllustrisTNG-ODM simulation as the projection plane. We have verified that our results remain unchanged when using the $x-z$ or $y-z$ planes. We select halo pairs on the 2D plane, requiring them to have halo masses in the same range as those analyzed in 3D. Additionally, these pairs must have projected separations in the range of $0 < d^{\mathrm{proj}}_{\mathrm{sep}} < 2~h^{-1} \mathrm{Mpc}$ and velocity differences $d_{\text{v}} < 1000~\mathrm{km/s}$ along the line of sight (LoS), specifically the $z$-axis. We account for both the position difference and the peculiar velocity difference when calculating the velocity difference $d_{\text{v}}$ between two given halos. Similar to the halo pairs selected in 3D, each of the projected halo pairs is additionally required to have no third halo within a cylinder centered at the midpoint of the halo pair, which has a projected radius equal to $d^{\mathrm{proj}}_{\mathrm{sep}}$ and a LoS depth of $\pm 1000~\mathrm{km/s}$. All subhalos located within $0.5d^{\mathrm{proj}}_{\mathrm{sep}}$ in projection and $\pm 500~\mathrm{km/s}$ along LoS from the nearer halo are considered satellites around the halo pair. For each satellite, we calculate the angle in the projection plane, $\Theta$, defined as the angle between the line connecting the satellite to the nearer host halo and the line connecting the two halos. Similar to the case in 3D, a satellite is labeled as being located ``inside'' the halo pair if $\Theta \in (0^{\circ}, 90^{\circ})$ and as being ``outside'' if $\Theta \in (90.0^{\circ}, 180.0^{\circ})$.

For the analysis of observational selection effects, we have constructed a mock catalog using the approach described in detail in Paper I. In short, assuming each subhalo in the simulation hosts a galaxy, we assign a stellar mass $M_\ast$ and an $r$-band absolute magnitude $M_r$ to each galaxy by applying the subhalo abundance matching model and the $M_\ast$ versus $M_r$ relation derived from the SDSS. We then use the model galaxies to construct the mock catalog, which incorporates the same selection effects as the SDSS galaxy sample, including redshift-dependent incompleteness due to the limiting magnitude and sky position-dependent incompleteness across the survey footprint, among other effects. From the mock catalog, we identify pairs of central galaxies as proxies for halo pairs and measure the angular distribution of satellite galaxies around each halo pair in the same way as described above in our 2D analysis. As shown in Paper I, the mock catalog successfully reproduces the observational measurements of $P(\Theta)$ for halo pairs of varying pair separations, primary halo masses, and secondary-to-primary halo mass ratios, as obtained from the SDSS galaxy sample.

In~\autoref{fig:3d_2d_mock_dms_pair} and~\autoref{fig:3d_2d_mock_dmr_pair}, we compare the satellite distribution around halo pairs as measured by $P(\Theta)$ on the projection plane (red lines) and in the mock catalog (blue symbols), for halo pairs with different separations, halo masses, and halo mass ratios. Note that the two lowest mass bins from previous figures have been combined into a single mass bin, $11.6 < \mathrm{log}{10}(M{\mathrm{p}}[\mathrm{M}_{\odot}/h]) < 12.6$, in order to avoid the relatively large Poisson noise present in those narrow bins in the mock catalog. As can be seen, the satellites selected in 2D and those from the mock catalog display very similar distributions around halo pairs with relatively small separations ($d_{\text{sep}}^{\text{proj}} < 1~h^{-1} \text{Mpc}$). At larger separations, however, the satellite distributions in 2D are slightly less bulging and lopsided than those in the mock catalog, with the discrepancy appearing to increase as $d_{\text{sep}}^{\text{proj}}$ increases. The relatively small differences found between the measurements from 2D and the mock catalog indicate that the 3D-to-2D projection is the dominant effect in a redshift survey like SDSS, and selection effects other than projection should not introduce significant biases, provided they are properly accounted for in the statistical measurement, as we have done in both Paper I and this work.

For comparison, the 3D distribution measured by $P(\cos\theta)$ from the previous section is plotted as the grey-shaded region in each panel. As expected, the 3D distribution is generally more strongly bulging than that measured in 2D and the mock catalog, particularly at large pair separations, low halo masses, and large halo mass ratios. For instance, as seen in the top-rightmost panel of ~\autoref{fig:3d_2d_mock_dms_pair}, the strongest bulging signal found in 3D for the most widely separated pairs of the lowest mass halos is dramatically suppressed by the projection effect, resulting in much weaker signals in the 2D plane and mock catalog. As a result, when transitioning from 3D to 2D and redshift space, the strong positive correlation of the bulging signal with pair separation is weakened, and the anti-correlation of the signal with halo mass is even reversed, leading to stronger bulging distributions around pairs of more massive halos. Furthermore, we notice that for pairs with the smallest separations ($d_{\text{sep}}^{\text{proj}} < 0.5~h^{-1} \text{Mpc}$), the tendency for a higher abundance of satellites on the outer sides of the halo pairs is also reversed by the projection effect.

One may wonder whether the differences observed between the 3D and 2D measurements arise not only from the projection effect but also from the different methods used to identify the halo pairs. To investigate this, we measure the distribution of the projected angle, i.e., $P(\Theta)$, for the halo pairs selected in 3D, and we present these measurements as green symbols in \autoref{fig:3d_2d_mock_dms_pair} and \autoref{fig:3d_2d_mock_dmr_pair}. As can be seen, these measurements agree very well with those from the 2D analysis (the red-shaded regions), demonstrating that the different means of constructing halo pair samples contribute little to the differences observed between the 3D and 2D measurements.

\subsection{Satellite distribution around overlapping halos}

Now we further examine the influence of projection and observational selection effects on the overlap effect. Similar to the 3D analysis, we construct samples of overlapping halos with and without incorporating halo alignment, both in 2D and in the mock catalog. For the overlapping halos without alignment, two halos selected from the projection plane or the mock catalog are placed at the same projected separation and velocity difference as the corresponding halo pairs, and their halo masses are matched between the overlapping halos and the halos of the pair, using the same tolerance as adopted in the 3D analysis. For the overlapping halos that include alignment, we estimate the major axis of each halo in the projection plane from the projected inertia tensor, which is calculated from the second moment of mass derived from the projected particle distribution. The ``aligned overlapping samples'' in the projection plane and the mock catalog are then constructed by rotating the major axes of the overlapping halos, along with their surrounding satellite distribution, so that the major axes of each pair of halos are aligned along their connection line.

\begin{figure*}[ht!]
    \centering
	\includegraphics[width=0.8\textwidth]{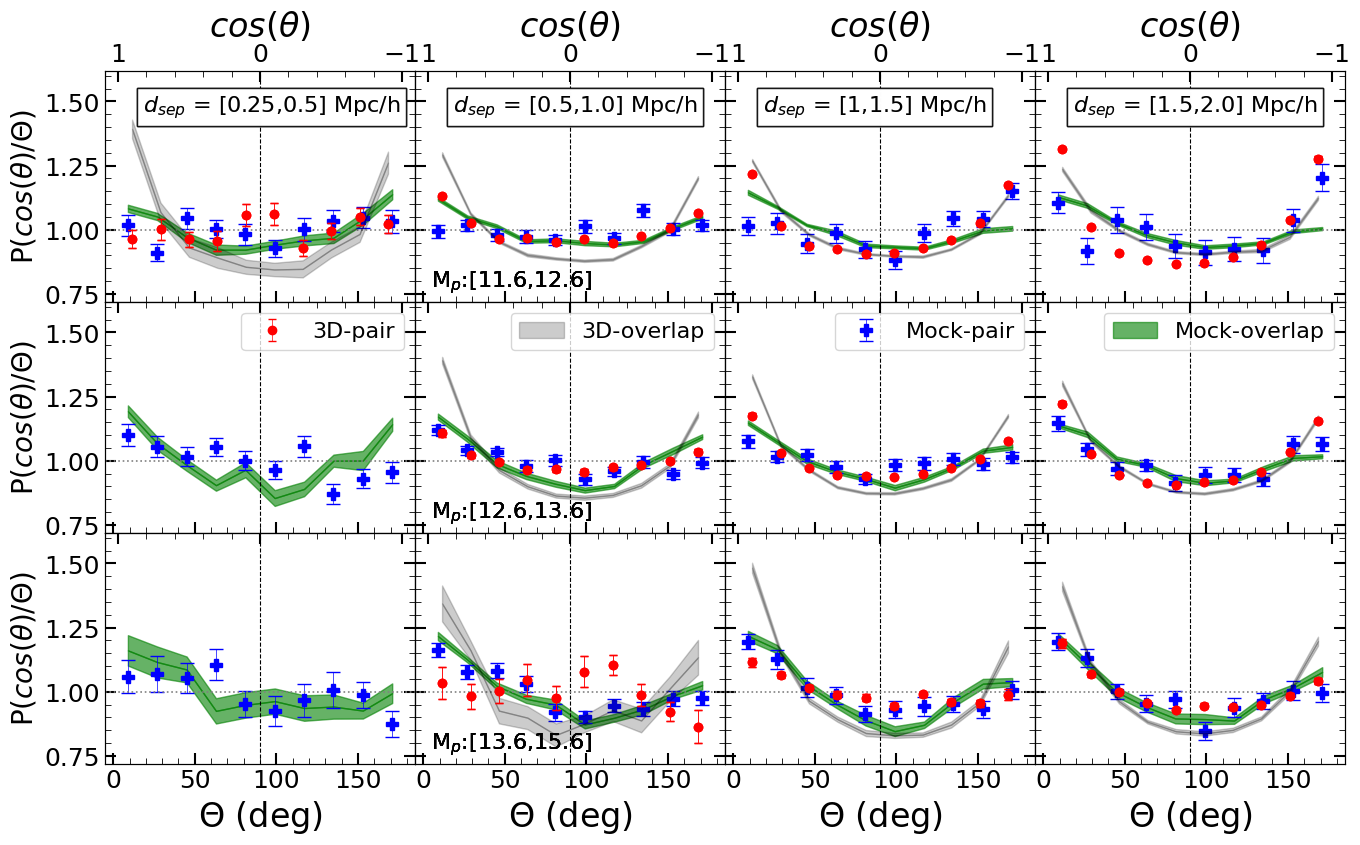}
    \caption{Measurements of satellite distribution are compared for actual halos in 3D (red), overlapping halos in 3D (grey), actual pairs in the mock catalog (blue) and overlapping halos in the mock catalog (green). Panels from left to right are for different pair separations, and those from top to bottom are for different primary halo masses, as indicated.}
    \label{fig:3d_mock_pair_overlap_1}
\end{figure*}

\begin{figure*}[ht!]
    \centering
	\includegraphics[width=0.8\textwidth]{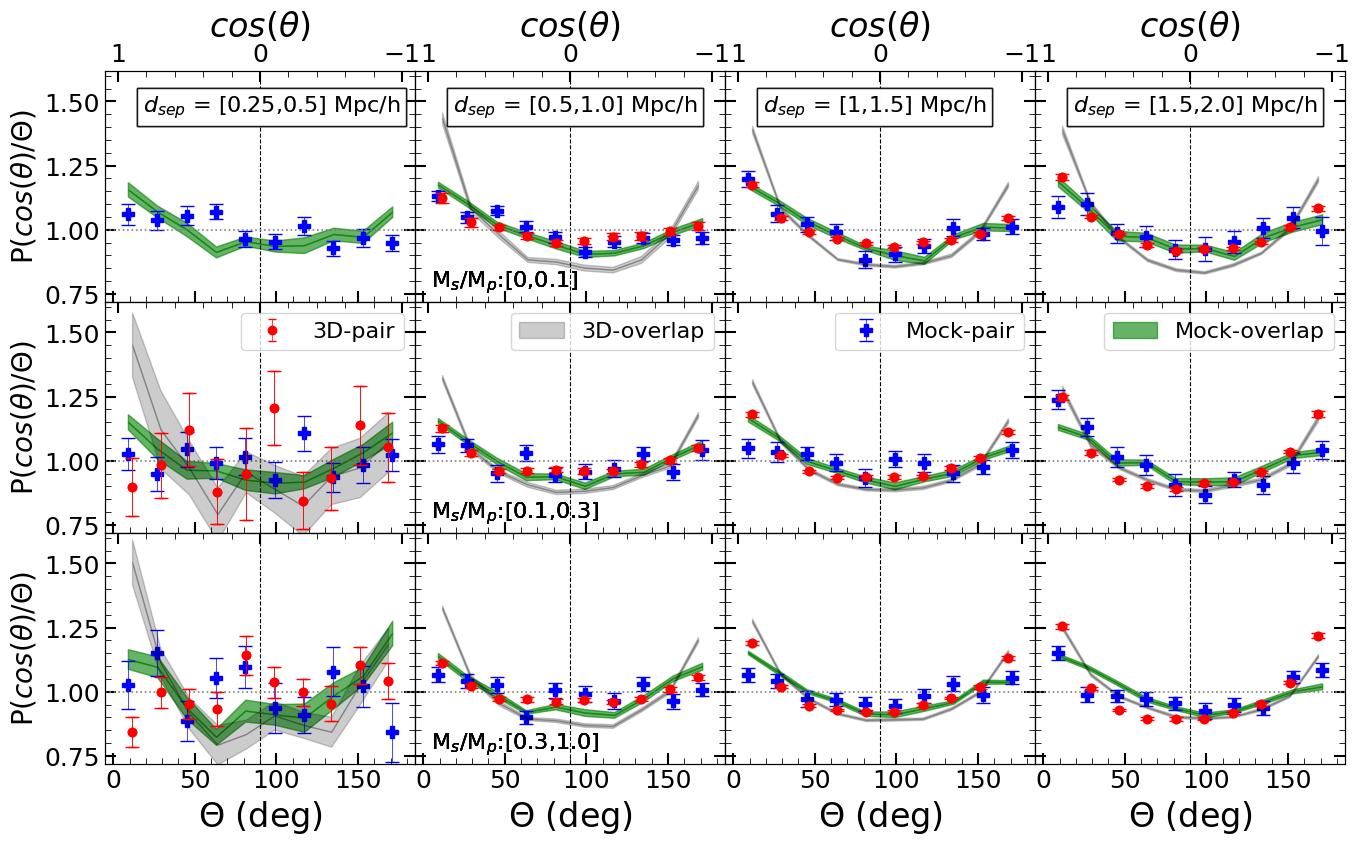}
    \caption{Same as the previous figure, but panels from top to bottom are for different secondary-to-primary halo mass ratios, as indicated.}
    \label{fig:3d_mock_pair_overlap_2}
\end{figure*}

For the overlapping halos without alignment, we find that the satellite distribution quantified by $P(\Theta)$ exhibits no significant differences among 3D, 2D, and the mock catalog for halo pairs with varying separations, halo masses, and mass ratios. For the aligned overlap samples, we observe that the bulging distribution found in 3D is significantly weakened in both 2D and the mock catalog due to the projection effect. In \autoref{fig:3d_mock_pair_overlap_1} and \autoref{fig:3d_mock_pair_overlap_2}, the shaded regions in grey and green show the measurements of $P(\Theta)$ for the aligned overlapping halos in 3D and the mock catalogs. For clarity, we do not display the 2D measurements, as they are very close to those from the mock catalog. As can be seen, the weakening due to projection occurs by a factor that appears to be larger at smaller separations and for overlapping halo pairs with more massive halos and smaller mass ratios. As a result, the satellite distribution in the overlap samples from the mock catalog shows a rather weak dependence on halo pair properties, unlike the distribution in 3D, which exhibits strong variation in the bulging signal from panel to panel, as illustrated in the figures.

For comparison, the results for the actual halo pairs, as obtained earlier in 3D (red symbols) and from the mock catalog (blue symbols), are additionally shown in \autoref{fig:3d_mock_pair_overlap_1} and \autoref{fig:3d_mock_pair_overlap_2}. In the mock catalog, the actual halo pairs and the overlapping halos demonstrate good agreement in satellite distribution across different halo pair properties. However, this agreement is coincidental and arises from the projection effect, which weakens the bulging signal in different ways for the actual pairs and the overlapping halos. Therefore, caution should be exercised when interpreting the observational distribution of satellites in halo pairs through the lens of the overlap effect.

\section{Discussion}
\label{sec:discussion}

\subsection{Comparison with previous works}
\label{sec:3_3}

Previous studies of satellite distribution have focused on the MW-M31 system in the Local Group \citep{2006McConnachie, 2013Conn, 2013Ibata}, as well as its analogs in the nearby Universe \citep{2016Libeskind} and in cosmological simulations \citep{2017Pawlowski, 2019Gong}. The lopsided and bulging distribution of satellites observed in the Local Group is similarly observed for pairs of galaxies in the SDSS that broadly resemble the MW-M31 system \citep{2016Libeskind}. Additionally, cosmological simulations present similar satellite distributions around analogous halo pairs, demonstrating that the observational results are not purely caused by projection effects \citep{2017Pawlowski, 2019Gong}. Furthermore, by placing isolated halos at the same pair separations as the actual halo pairs, these studies have shown that the lopsidedness and bulging of satellites in the Local Group and analogous systems cannot be simply attributed to the overlap effect. 

Generally, our work yields results that are broadly consistent with those of previous studies for cases analogous to the MW-M31 system, specifically for the subsample of halo pairs with halo mass $11.6\leq\log_{10} (M_{\mathrm{p}}/\mathrm{M}_{\odot})\leq 12.0$ and pair separation $1.0\leq d_{\mathrm{sep}}/(h^{-1}\mathrm{Mpc}) \leq 1.5$. The trends of bulging and lopsidedness signals with primary halo mass, halo mass ratio, pair separation, and satellite-to-host distance, as identified in our work, are consistent with those reported by \citet{2019Gong}. In addition, we have performed the same test on the overlap effect as conducted in previous studies, and we also find that the satellite distribution around actual halo pairs cannot be explained by the overlap effect if the alignment of halo orientation is not taken into account, as was the case in previous investigations.

In this work, we have attempted to extend previous studies in several key aspects. Firstly, we consider more general cases of halo pairs beyond the Local Group by examining halo pairs with wider ranges of halo mass and pair separation. We find that the lopsided and bulging distributions are not limited to the MW-M31 analogs but are observed for halo pairs across broad ranges of halo mass and pair separation. Secondly, for the first time, we take into account the spatial alignment of halos when testing the overlap effect. We find that although the lopsidedness remains similar, the inclusion of halo alignment produces significant signals of bulging, thereby naturally reproducing the satellite distribution for halo pairs with varying properties. Additionally, we conduct comparisons between analyses conducted in 3D and 2D, as well as in the mock catalog, thus providing a comprehensive examination of the effect of projection and selection effects in observational samples.

\subsection{Implications for the origin of bulging and lopsidedness}
\label{sec:discussion_result}

Our comparison between the 3D and 2D analyses demonstrates that, although the projection effect weakens the bulging distribution by a factor that depends on pair separation and halo mass, the bulging and lopsided distributions of satellites are fundamentally intrinsic behaviors of the halo pairs in a $\Lambda$CDM universe. More importantly, we find that the satellite distribution around halo pairs with varying properties can be reasonably well reproduced by the overlap effect, either with or without including the spatial alignment of halos. Specifically, the overlapping halos that do not consider halo alignment cannot produce any bulging signal, but they successfully replicate the satellite distribution for halo pairs that are closely separated or involve massive halos. In contrast, the overlapping halos that are aligned with each other can produce significant bulging distributions that match the satellite distribution around halo pairs with relatively large separations and low halo masses. 

As noted at the end of \autoref{sec:3danalysis}, these results imply that the lopsidedness and bulging of the satellite distribution may arise from two distinct origins: the alignment of halo orientation with large-scale filaments is responsible for the bulging, while the overlap effect without alignment accounts for the lopsidedness. It is likely that the lopsidedness produced by the overlap effect is a ubiquitous phenomenon applicable to all halo pairs, exhibiting weak dependence on pair properties. In contrast, the bulging distribution generated by halo alignment is more dependent on the specific characteristics of the pairs. As a combined result of these two effects, the overall satellite distribution thus demonstrates varying strengths of lopsidedness and bulging across halo pairs with different properties. 

The excess of matter between dark matter halo pairs has been well established in both simulations and observations, and this excess has indeed been utilized as a common method to identify filaments \citep{2005Colberg, 2016Clampitt, 2017Epps, 2020Xia, 2020Kondo, 2022Yang}. Additionally, previous studies have found that galaxy pairs \citep{2015Tempel, 2018Mesa, 2024Sarkar} and galaxy multiplets composed of small groups of 2-4 member galaxies \citep{2024Rong, 2024Lamman} tend to align with their host filaments, with stronger alignment observed for more widely separated halo pairs \citep[e.g.][]{2015Tempel}. Moreover, the thickness of filaments is found to be approximately 1-2 $\text{Mpc}$ in the local universe \citep[e.g.][]{2010Arag, 2010Bond, 2014Cautun, 2024Wang}, which is consistent with our finding that the bulging signal dominates over the lopsidedness for wide pairs with $d_{\text{sep}} \gtrsim 1~h^{-1} \text{Mpc}$. All of these results align well with the idea that the bulging distribution has an alignment origin, as described above. This picture is also consistent with our finding that the bulging distribution is primarily contributed by subhalos that are relatively distant from their host halos, considering that the satellites in the outskirts of halos are more recently accreted and thus more capable of keeping their anisotropic distribution inherited from the filaments (\citealt{2013Li}; \citealt{2019Gong}). 

One may wonder whether and how the observed lopsided and bulging satellite distributions around halo pairs relate to those found around isolated halos/galaxies. Previous studies attribute both anisotropic distributions around isolated systems to large-scale tidal fields and anisotropic accretion (references in~\autoref{sec:intro}). Our analysis reveals a distinct physical origin for lopsidedness in paired systems: it arises primarily from geometric overlap effects due to halo alignment, rather than contributions from isolated-halo lopsidedness or local gravitational mechanisms. This contrasts with the anisotropic accretion-driven lopsidedness found in isolated galaxies \citep[e.g.][]{2024Liu}. For the bulging distribution, we find that satellite alignment with large-scale filaments—likely tied to the orientation of paired halos—provides a plausible explanation. This mechanism resembles the origin of bulging distributions around isolated halos, whose orientations also tend to align with the large-scale dark matter distribution \citep[e.g.][]{2008Faltenbacher}. However, a key difference emerges in the mass dependence: while isolated halo alignment strengthens with increasing halo mass, the bulging signal in paired systems intensifies for lower-mass pairs at fixed separation. This inverse mass trend has two implications. First, low-mass halo pairs trace large-scale filaments more effectively than massive pairs, which preferentially reside in high-density knots connecting multiple filaments \citep[e.g.][]{2010Arag, 2018Codis, 2021Gouin}. Second, the contrasting mass dependencies may indicate distinct physical origins for bulging in paired versus isolated systems. While our model of aligned overlapping halos broadly explains the satellite distribution around halo pairs of various properties, discrepancies remain unresolved in certain cases. A direct, systematic comparison of satellite anisotropy between isolated and paired systems would be  essential to fully understand their distribution mechanisms in both contexts.


\subsection{Projection and selection effect}

We have examined the effect of projection and observational sample selection on the satellite distribution in halo pairs by comparing the results obtained in 3D, 2D, and the mock catalog. The consistency between the 2D analysis and the mock catalog demonstrates that selection effects other than projection should not introduce significant biases, provided that the measurements are statistically corrected for those effects, as is done in both Paper I and this work. The 3D-to-2D projection significantly suppresses the bulging signals by a factor that depends on halo pair properties. Specifically, the suppression is particularly strong for widely separated pairs of low-mass halos. As a result, the anti-correlation of bulging with halo mass found in 3D turns into a positive correlation in the observational sample. 

In addition, the 3D-to-2D projection affects the satellite distribution around actual halo pairs and that around overlapping halos in different ways. The projection effects exert a significantly stronger influence on overlapping systems compared to the pairs. This disparity primarily arises because the projection distance exceeds the intrinsic 3D separation. On the one hand, the bulging signal is stronger for the wider-separated pair system, thereby eliminating some projected effects from the interiors of the satellites. On the other hand, the bulging signal is further weakened for overlaps due to the weaker bulging feature for larger separation of the overlaps. For instance, as can be seen from \autoref{fig:3d_mock_pair_overlap_1}, although the overlap sample and the actual pair sample agree well with each other when analyzed in 3D, their 2D measurements obtained with the mock catalog present significant discrepancies, particularly for widely-separated pairs of low-mass halos. This echos the similar discrepancies found between halo pairs and overlapping halos from the SDSS galaxy sample; see Fig. 11 and discussion at the end of Section 4 in Paper I.

Therefore, caution should be exercised when interpreting the observational measurements of satellite distribution. The impact of projection cannot be easily corrected during statistical measurements, and a correct understanding can be achieved only by carrying out careful comparisons between model and data with the assistance of mock catalogs. 

\section{Summary}
\label{sec:conclusions}

In this work, we investigate the distribution of subhalos (satellites) in the vicinity of dark matter halo pairs within cosmological simulations. We utilize the Illustris-TNG300 dark matter-only simulation, from which we select a sample of halo pairs covering a wide range of primary halo mass, secondary-to-primary halo mass ratio, and pair separation. We first perform three-dimensional (3D) measurements of the satellite distribution around the halo pairs, examining the dependence on pair separation, halo mass, mass ratio, and satellite-to-host distance. Next, we analyze the overlap effect by constructing samples of overlapping halos. In particular, we include the spatial alignment of halos in the overlap effect by aligning the major axes of the overlapping halos. Finally, we investigate the impact of projection and observational selection effects by conducting the same analysis of satellite distribution in two dimensions (2D) and in a mock catalog constructed from the same simulation, which replicates the selection effects of the SDSS galaxy sample as used in our Paper I (\citealt{2025Guo}).

Our main findings can be summarized as follows. 

\begin{enumerate}
    \item The satellites around halo pairs with separations $d_\text{sep}>0.5~h^{-1}\text{Mpc}$ present a non-uniform distribution in 3D. This distribution is a combined result of two distinct features: the ``bulging'' distribution, characterized by an overabundance along the connection line of the paired halos, and the ``lopsided'' distribution, which shows an overabundance in the region between the paired halos relative to the regions on the outer sides.
    \item The bulging signal is stronger for halo pairs that are more widely separated and involve less massive halos, but it exhibits a rather weak dependence on the secondary-to-primary halo mass ratio. This signal is primarily contributed by subhalos that are relatively distant from their host halos.
    \item Compared to the bulging, the lopsidedness shows weaker dependence on pair properties. Nevertheless, the lopsidedness strengthens as the primary halo mass increases, but it only weakly depends on pair separation. This signal also exhibits a weak dependence on the halo mass ratio and is primarily attributed to more distant subhalos. 
    \item The subhalos around close pairs with $d_\text{sep}<0.5~h^{-1}\text{Mpc}$ appear to exhibit relatively uniform distributions, or even slightly higher abundances on the outer sides, though with large uncertainties. 
    \item The satellite distribution around halo pairs with varying properties can be reasonably reproduced by the overlap effect, provided that the spatial alignment of halos is properly taken into account. Specifically, overlapping halos without alignment can replicate the satellite distribution for halo pairs that are closely separated or involve massive halos, while overlapping halos that are aligned with each other can produce strong bulging distributions that match the satellite distribution around halos pairs with large separations and low-mass halos.
    \item Selection effects other than projection in real galaxy samples introduce negligible biases in the measurements of satellite distribution, but 3D-to-2D projection significantly suppresses the bulging signal, with particularly strong effect at large pair separations, low halo masses, and large halo mass ratios. This projection effect cannot be easily corrected, and a correct understanding can be achieved only by carrying out careful comparisons between model and data with the help of mock catalog.  
\end{enumerate}

\section*{Acknowledgments}.
This work is supported by the National Key R\&D Program of China (grant NO. 2022YFA1602902), the National Natural Science Foundation of China (grant Nos. 12433003, 11821303, 11973030), and the China Manned Space Program with grant no. CMS-CSST-2025-A10.

\bibliography{sample631}{}
\bibliographystyle{aasjournal}



\end{document}